# Introducing new resonant soft x-ray scattering capability in SSRL

Cheng-Tai Kuo,[1,a)] Makoto Hashimoto,[1] Heemin Lee,[1] Tan Thanh Huynh,[1] Abraham Maciel,[1] Zina Zhang,[1,2] Dehong Zhang,[1] Benjamin Edwards,[3] Farzan Kazemifar,[3] Chi-Chang Kao,[1,4] Donghui Lu,[1] and Jun-Sik Lee[1,b)]

[1]*Stanford Synchrotron Radiation Lightsource, SLAC National Accelerator Laboratory, Menlo Park, California 94025, USA*

[2]*University of California, Davis, California 95616, USA*

[3]*Department of Mechanical Engineering, San Jose State University, Sam Jose, California 95192, USA*

[4]*SLAC National Accelerator Laboratory, Menlo Park, California 94025, USA*

Resonant soft X-ray scattering (RSXS) is a powerful technique for probing both spatial and electronic structures within solid-state systems. We present a newly developed RSXS capability at beamline 13-3 of the Stanford Synchrotron Radiation Lightsource (SSRL), designed to enhance materials science research. This advanced setup achieves a base sample temperature as low as 9.8 K combined with extensive angular motions (azimuthal $\phi$ and flipping $\chi$), enabling comprehensive exploration of reciprocal space. Two types of detectors—an Au/GaAsP Schottky photodiode and a CCD detector with over 95% quantum efficiency—are integrated to effectively capture scattered photons. Extensive testing has confirmed the enhanced functionality of this RSXS setup, including its temperature and angular performance. The versatility and effectiveness of the system have been demonstrated through studies of various materials, including superlattice heterostructures and high-temperature superconductors.

[a)] Author to whom correspondence should be addressed. Electronic mail: ctkuo@slac.stanford.edu

[b)] Author to whom correspondence should be addressed. Electronic mail: jslee@slac.stanford.edu

## I. INTRODUCTION

Exploring the interplay of degrees of freedom, such as charge, spin, orbit, and lattice, within solid materials is crucial for understanding their functionality. Many emerging functionalities have been reported in complex materials, commonly referred to as strongly correlated systems, including 3*d* transition metal oxides and rare-earth compounds, which exhibit phenomena such as (anti-) ferromagnetism, multiferroicity, superconductivity, and two-dimensional topological properties. Resonant soft X-ray scattering (RSXS) has been appreciated as a valuable tool for probing these degrees of freedom in correlated systems, particularly when they are spatially modulated [1-7]. The nature of soft X-rays also enables exploration not only of spatially ordered degrees of freedom but also of their electronic structure through resonant effects via strong dipolar transitions, particularly in transition metal *L* edges, rare earth *M* edges and oxygen *K* edges [5].

In 2015, the first RSXS setup at Stanford Synchrotron Radiation Lightsource (SSRL) was launched at beamline 13-3, like other facilities around the world [8-16]. The setup featured a kappa-type in-vacuum diffractometer, with limited azimuthal ($\phi \sim \pm 5°$) and flipping ($\chi \sim \pm 5°$) angles. The sample was cooled using a copper braid connected to an external liquid helium cryostat, achieving a base temperature of approximately 23.5 K due to the intrinsic thermal path to the room temperature of the diffractometer. This setup successfully supported many RSXS-dedicated research efforts until 2022 [17-33]. Despite these successes, SSRL recognized the need for a new RSXS setup capable of reaching even lower temperatures with a wider angular range. This is critical, as many interesting quantum phenomena occur below the current base temperature. For instance, in our previous study of the high-$T_c$ cuprates, $La_{2-x}Sr_xCuO_4$ (LSCO) [23], we observed complex interactions between charge density wave (CDW), spin density wave (SDW), and superconductivity (SC) at around 25 K – 30 K. However, the limited base temperature of the RSXS setup prevented a more thorough exploration of these effects.

To overcome these limitations, SSRL initiated a significant design effort to develop a new RSXS setup (see Fig. 1a). This new setup is designed to achieve lower temperatures and provide greater coverage in reciprocal space and higher accuracy. First, we adopted a manipulator-type concept rather than a goniometer type, which offers a simpler thermal path between the sample and the cryostat cold head. This design also allowed us to implement a proper radiation shield around the sample area, including a sample cover door operated by an in-vacuum motor (Fig. 2b). Second, we included two additional in-vacuum motors dedicated to controlling the $\phi$ and $\chi$ angles, expanding the angular range in reciprocal space. Third, this design was optimized using Computational Fluid Dynamics (CFD) and Finite Element Analysis (FEA) simulations, allowing us to fine-tune the components. The newly developed RSXS setup at SSRL can now reach a base temperature of 9.8 K while maintaining a wide range of $\chi$ (from +30° to -30°) and $\phi$ (from +90° to -90°) motions. This enhanced capability will enable X-ray experiments to explore previously hidden phases and phenomena near the ground state of complex materials and novel magnetic systems.

The following sections will describe in more detail the overall design and performance of our new RSXS setup. Section II provides an overview of the design of the diffractometer components. Section III focuses on the performance, particularly highlighting cooling mechanisms and thermodynamic CFD-FEA simulations as well as the motion components. Finally, Section IV presents experimental demonstrations, showcasing scattering studies on a superlattice heterostructure and a high-temperature superconductor.

## II. DESIGNS

### *a.* Motion components:

In developing the mechanical motion of a low-temperature RSXS instrument, two major considerations need to be addressed simultaneously, namely maintaining UHV environment while rotating and positioning the sample and detector, at the same time minimizing thermal loss while protecting the sample from radiation. To achieve these, we strategically designed the operational degrees of freedom, maximizing their placement outside the vacuum chamber. This significantly reduces out-gassing during motion and facilitates easy access, enhancing overall performance and maintenance.

As shown in Fig. 1b, two differentially pumped rotating platforms (DPRP; McAllister Technical Services) positioned at the top and bottom facilitate sample ($\theta$) and detector ($2\theta$) rotating motions. The top and bottom DPRPs are equipped with independent *x*-*y* (Gon-*x*/*y*) and *z* (Gon-*z*) motion manipulators (McAllister Technical Services), ensuring alignment of the rotation centers. These aligned centers can be realigned with the X-ray beam direction using a vertically movable chamber supporting table, see Fig. 1a. Additionally, an *x*-*y*-*z* manipulator (Sam-*x*/*y*/*z*) allows for independent adjustment of the sample position relative to the beam.

To minimize thermal gradient and footprint, two angular motions — sample flipping ($\chi$) and azimuthal ($\phi$)—are operated using in-vacuum motors (Arun Microelectronics Ltd.) inside the chamber. This design not only ensures ultra-high vacuum (UHV) conditions but also enhances motion accuracy under a minimized footprint (see Fig. 2a). Additionally, the installation of a motorized shielding door with a 0.3-inch diameter hole around the sample holder provides not only connivence of sample transfer but also achieves further improvement of thermal loss reduction, as shown in Fig. 2b. The detailed information of the linear and angular motions is summarized in Table I. Despite these in-vacuum motions, the developed endstation achieved an ultimate base pressure of $1 \times 10^{-10}$ Torr, providing an additional benefit for exploring surface-sensitive systems.

### *b.* Detector components:

Figure 3a shows the inside view of the main chamber, where the scattering plane of the diffractometer is horizontal. The detector ($2\theta$) stage, connected via the bottom DPRP and $z$-motion manipulator, hosts three components. The first is a pre-aligned slit structure at the 180-degree mirror positions, facilitating quick alignment checks of the detector center against the X-ray beam. The second component consists of 2 point detectors with different pinhole sizes (1 mm and 2 mm diameters), using an Au/GaAsP Schottky photodiode (PD, Hamamatsu Photonics) to improve low-photon sensitivity. The third component is a charge-coupled device (CCD) area detector (GE-VAC, Greateyes GmbH), featuring 256 (horizontal) × 1024 (vertical) pixels, each measuring 26 μm × 26 μm. In our horizontal scattering geometry, the CCD is positioned approximately 400 mm away from the sample, covering a vertical range of approximately 7 degrees and a horizontal range of 1.75 degrees. To maximize quantum efficiency, a proportional–integral–derivative (PID)-controlled chilled-water-cooling system is connected to the CCD structure. Notably, there is no interference when switching between detectors (PD or CCD), as they are separated by 240 degrees on the single $2\theta$ stage and rotate together, as shown in Fig. 3b. The direction of vertically moving support table with respect to the incoming X-ray is also shown in Fig. 1a. Finally, it is worth noting that the diffractometer was carefully designed to prevent interference among the in-vacuum components, including the sample manipulators, detector cooling lines, and cables. The solution we developed allows smooth movement of the detector and its related components without restricting motor motion or causing interference inside the chamber.

### *c.* Cooling components:

Given the complex structures, including the in-vacuum motions, it was critical to design a method to thermally isolate the sample position and other structures. The core engineering concept comprised two key aspects – (i) maximizing cooling of the sample stage using bundles of copper braid attached from the cryostat cold head (Janis Cryostat ST-400, Lake Shore Cryotronics), and (ii) minimizing the thermal losses to the mechanical parts, including the in-vacuum motors. For the first aspect, a radiation shielding structure made of copper and bronze, positioned close to the sample stage, serves as a thermal radiation barrier and a cooling reservoir (see Fig. 4a). For the second aspect, we have added thermal breakers (VESPEL SP1/SP3, brown colored in Fig. 4a) at critical locations (e.g., connections to in-vacuum motors and other support structures) that would otherwise cause significant thermal losses and increase the minimum achievable sample temperature. This strategy enables the creation of a long thermal path and maintains larger temperature differences between the sample and surrounding components. The cooling performance of the designed structure, particularly the sample temperature, was evaluated a priori through CFD and FEA simulations. Details of the simulation can be found in Ref. 34.

## III. PERFORMANCES

### *a.* Cooling components:

For evaluation of the cooling performance of the designed structure, we mostly focused on tracking temperature behaviors on three locations: cold head, shield side, and the sample position. Figure 4b shows the simulations of cooling (lines) and the corresponding experimental results (open circles). The cooling experiment was carried out using an open-cycle liquid helium dewar. As expected from the simulations, there is large temperature difference compared to the purposely cooled interior bodies. In the simulations, despite having little temperature variation across the thin shield bodies, the overall shields area shows a moderate cooling performance and reaches approximately 118.6 K, demonstrating good agreement with the experimental value of 120 K. On the other hand, due to the well isolated thermal shield, the sample settles down at the ultimate simulated temperature about 12 K, which is close to the experimental value of 11.3 K. The shield and the sample temperatures show a clear contrast within the manipulator head, indicating that the VESPEL thermal breakers are functioning effectively. Although the overall experimental results match well with simulations, we found that the base temperature is highly sensitive to the shield temperature. And, by modifying the design of cooling paths to the shield and sample, we were able to lower the base temperature of the sample stage to 9.8 K (see the open downward triangle in the inset of the bottom figure in Fig. 4b).

To ensure precise temperature control during measurements, three silicon-diode sensors (Lake Shore Cryotronics, DT-670SD) are placed at key locations: the sample, the cold head (liquid helium injection point), and the shield. The sample sensor is positioned less than 1 cm behind the sample, with thermally anchored leads to minimize errors. A temperature controller (Lake Shore Cryotronics, Model 336) employs a four-lead technique for precision. The sample stage and holder, made of high-thermal-conductivity materials (OFHC copper for the stage/holder, and either aluminum or OFHC copper for the sample post), enhance thermal coupling. While a potential thermal gradient may exist due to sensor placement, measurements confirm accuracy down to 17 K, with a verified offset less than 1 K between the sample temperature reading and the known superconducting transition temperature ($T_c$). Also, it is important to note that further assessment below 17 K is needed to evaluate any temperature discrepancies.

Additionally, a closed-cycle helium liquefied system (Janis RGC4, Lake Shore Cryotronics, see Fig. 1a) was installed to cool down the diffractometer. The entire system reached a thermally stable condition approximately 6 to 7 hours after the initial cooling from room temperature. The ultimate temperature of the sample stage stabilizes at ~11 K via fine adjusting the needle valve position of the RGC4 cryostat. Finally, it is noteworthy that there is no significant difference in thermal performance between the open and closed-cycle helium systems (i.e., liquid helium dewar vs. RGC4 system).

***b.* Motion components:**

The critical angular motions for X-ray scattering involve four circles: sample ($\theta$), detector ($2\theta$), azimuthal ($\phi$), and flip ($\chi$) motions. Testing of each motion of the new diffractometer was conducted through visual inspection (via multiple optical cameras), specular reflectivity, and structural Bragg reflection.

With the conventional specular X-ray reflectivity (XRR), we checked the rotation center between $\theta$ and $2\theta$. For this test, we employed a superlattice film, $[SrRuO_3$(m unit cell)/$SrTiO_3$ (n unit cell)$]_{10}$ (SRO/STO), grown on a $DyScO_3$ single crystal substrate. This is because it is expected to generate many Kiessig fringes, and, in particular, the specular condition is sensitive to the rotation center alignment. In addition, the incident photon energy $h\nu$ = 1500 eV for this XRR test was intentionally tuned to increase the $q_z = 4\times\sin(2\theta/2)/\lambda$ range and see as many Kiessig fringes as possible. Figure 5a shows a typical reflectivity pattern of a superlattice structure. Note that the scattering X-rays were collected by a PD, which is one of our detector options. The high-frequent oscillation part in the Kiessig fringes is attributed from the total thickness, which is well matched with the predicted 10 times bilayer SRO/STO. The strong peak per every 9 oscillation originates from the interference effect of the repeated bilayers. Beyond the detailed discussion of the XRR pattern, this XRR test's main purpose was for confirming the rotation center. After checking the specular condition at the low $\theta$–$2\theta$ area (about 10° – 20°), specular conditions were well produced over the full angular range until the beam is blocked by the detector around $2\theta$ = 156° (marked by an arrow in Fig. 5a). This demonstration indicates that the rotation centers between $\theta$ (sample) and $2\theta$ (detector) are well aligned.

In addition, we also checked the angular resolution through a crystal's structural Bragg reflection. Figure 5b shows a YBCO crystals' (0 0 2) reflection measured at $h\nu$ = 1777 eV. The peak is very sharp, indicating that our angular resolution is, at least, 0.02°. Based on the aligned rotation center position of the sample, we further monitored $\phi$ and $\chi$ motions using the optical cameras. For this test, we mounted a YAG crystal, which showed the illumination when X-rays shined on the sample position. As shown in Fig. 5c, we could confirm that the beam position on the sample is nearly identical within a few microns tolerance while moving $\phi$ and $\chi$ motions.

## IV. RSXS DEMONSTRACTION

### a. Detecting a magnetic pattern on superlattice

Exploring magnetic properties, such as momentum, spin direction, ordering pattern, etc., is crucial for utilizing these properties in real-world applications. While several techniques can estimate magnetic moment (mostly transport measurements), X-ray and neutron scattering uniquely offer detailed insights, particularly into spin ordering and configurations. Due to the element specific sensitivity of X-ray technique, resonant X-ray magnetic scattering has been gaining prominence for investigating both ferromagnetic and antiferromagnetic materials [1,5,35,36]. For example, when non-magnetic ions are directly bonded with ferromagnetic ions within a unit cell structure, these non-magnetic ions can also become ferromagnetically polarized. Although this induced polarization may be weak due to indirect exchange coupling, it can still be detected under specific conditions using resonant X-ray techniques. Additionally, lowering the temperature can enhance spin polarization, making it easier to detect these subtle effects. In this context, it would be valuable to test whether the newly developed scattering capabilities at SSRL can detect induced magnetic moments, particularly focusing on their spin configurations.

For this test, we employed an SRO/STO superlattice sample. A neodymium permanent magnet post (~0.4 Tesla) was placed under the sample that allows the application of an in-plane magnetic field relative to the scattering plane for this experiment. According to a previous work on $SrRuO_3$ thin film [24], it is expected that oxygens bonded with the ferromagnetic Ru cations could become ferromagnetically polarized, potentially forming a skyrmion pattern together with Ru. To explore this hypothesis, we firstly conducted oxygen *K*-edge X-ray magnetic circular dichroism (XMCD) in this scattering chamber. Note that the experimental configuration is shown in the inset of Fig. 6a, and all measurements were conducted at $T$~13 K. With varying two polarizations (RCP/LCP: right/left circular polarization) the XMCD signal at $h\nu$ ~ 527.5 eV, corresponding to a hybridization state of Ru and O, is pronounced. This indicates that the oxygen atoms are ferromagnetically polarized by the Ru.

As a next step, we conducted the magnetic reflectivity measurements with RCP and LCP at $h\nu$ ~ 527.8 eV, as shown in Fig. 6b. A clear magnetic asymmetry of oxygen atoms along the $q_z = 4\times\sin(2\theta/2)/\lambda$ direction is observed and demonstrated. Note the magnetic asymmetry ratio is determined by the equation $MAR = (I_{\mathrm{LCP}} - I_{\mathrm{RCP}})/(I_{\mathrm{LCP}}+I_{\mathrm{RCP}})$. In particularly, we found that the magnetic asymmetry ratio cancels out at the superlattice peak position, where the $SrRuO_3/SrTiO_3$ bilayer effect is reinforced. To explore the spin configuration, we adopted this effect – we set the scattering angles at this position and investigate the patterns with varying the incident polarizations. A difference pattern between two polarizations is shown in Fig. 6c, possibly indicating a rotating spin configuration, which is very similar with the previous work [24]. In addition, we utilized the capability of this new instrument to rotate the $\phi$-angle ($\phi = 30°$) to verify whether this pattern was an artifact. As shown in Fig. 6d, the magnetic pattern remains constant while the raw patterns rotate, suggesting that the spin configuration forms a rotating formation in the film plane. This observation potentially implies the existence of a skyrmion formation at the interface of the bilayer. Note that more detailed analysis is discussed elsewhere [37].

### b. Detecting charge-density wave in high-$T_c$ cuprates

High-temperature superconductors, particularly high-$T_c$ cuprates, have been the focus of extensive research since the discovery of their superconducting properties in 1986 [6,38]. The potential for zero-resistance conductivity in these materials could revolutionize modern energy grids. Consequently, researchers have conducted numerous studies, continually uncovering new phenomena that challenge existing paradigms of HTSCs, including the potential for room-temperature superconductivity. One such profound discovery is the ubiquitous presence of charge-density wave (CDW) states in high-$T_c$ cuprates [2,4,7,39,40]. For example, in La-based cuprates, a CDW order known as the stripe phase, with a periodicity of approximately four unit-cells ($q_{\text{cdw}} \sim 0.25$ *r.l.u.*), has been shown to correlate with the emergence of superconductivity [4,23,40,41].

To gain deeper insights into the nature of CDW order beyond its correlation with superconductivity, X-ray scattering techniques, particularly RSXS approach, have been actively employed in studies of high-$T_c$ cuprates [5,6]. This approach is advantageous because X-rays tuned to the Cu *L*-edge can directly detect modulations in copper within the superconducting $CuO_2$ planes of cuprates. Therefore, performing RSXS studies at temperatures below the superconducting transition temperature is particularly beneficial for understanding the nature of CDWs. As mentioned earlier in the Introduction Section, this capability is one of the reasons behind the development of new scattering techniques at SSRL.

In this context, we demonstrate a RSXS study on $La_{2-x}Sr_xCuO_4$ ($x = 0.115$, LSCO) across its $T_c$, which is approximately 27 K. In the previous RSXS studies on this cuprate, it has been reported that the stripe-formed CDW order continues to develop even below $T_c$ [23]. However, the previous RSXS setup at SSRL was not sufficient to investigate the detail regarding this development while temperature below $T_c$ because of limitation in the base temperature of the setup. In this regard, here, we newly present the CDW data measured at way lower than $T_c$. Figure 7a shows a schematic view of the RSXS measurement. Figure 7b shows the scattering map at *h* reciprocal space ~ 0.25 *r.l.u.* measured at 12 K (below $T_c$). In the projected intensity profiles (Fig. 7c) along the *h*-direction, one can observe that the CDW signal ( $-0.04 < k < 0.04$, blue line) is embedded with strong fluorescence background ($k < -0.04$, dot line). In the area $k < -0.04$, it can be regarded as pure fluorescence area. Figure 7d shows a scattering image after subtracting the fluorescence background, and the CDW signal is clearly observed. In the projected intensity profiles (Fig. 7e), it clearly indicates that the CDW signal persists at $T = 12$ K with respect to that at $T = 40$ K, meaning that the CDW correlation continues to strengthen below $T_c$. This finding suggests that there is more to the (hidden) interaction between CDW and superconductivity than a simple competition. Based on this demonstration, we confirm that the newly developed RSXS capability is functioning properly and holds the potential to explore new scientific opportunities at lower temperature regimes. Note that more detailed analysis is discussed elsewhere [42].

## V. CONCLUSION

We have developed a new resonant soft X-ray scattering (RSXS) endstation at beamline 13-3 of the Stanford Synchrotron Radiation Lightsource (SSRL), capable of achieving low temperatures and providing a wide range of angular motions. The new RSXS capability features a manipulator-type concept, in-vacuum motors for angular motions, and optimized cooling components, achieving a base temperature of 9.8 K. The performance of the new RSXS endstation was evaluated through visual inspection, X-ray reflectivity, and structural Bragg reflection, demonstrating its capability to probe the structural and electronic properties of complex materials at low temperatures. Preliminary scattering studies on a superlattice heterostructure and a high-temperature superconductor confirmed the effectiveness of the new RSXS endstation in exploring quantum phenomena. The new RSXS endstation will enable upcoming X-ray experiments to explore previously hidden phenomena near the ground state of material sciences.

## ACKNOWLEDGMENTS

The authors thank Prof. Masaki Fujita, Dr. Seunggyo Jeong, Prof. Deok-Yong Cho, and Prof. Woo Seok Choi for providing high-quality samples. All soft X-ray experiments were carried out at the SSRL (beamline 13-3), SLAC National Accelerator Laboratory, supported by the U.S. Department of Energy, Office of Science, Office of Basic Energy Sciences under Contract No. DE-AC02-76SF00515. FK acknowledges support from the U.S. Department of Energy, Office of Manufacturing and Energy Supply Chains under award no. DE-EE0010194.

| Function | Motor | Moving range | Note |
|---|---|---|---|
| diffractometer | $\theta$(th) | 0° to 180° | - |
| | $2\theta$(tth) | 0° to 167° | blocked angles: 156 ~164° |
| | $\chi$(chi) | 60° to 120° | normal case chi ~ 90° |
| | $\phi$(phi) | -90° to 90° | normal case phi ~ 0° |
| sample adjustment | samx (along x) | ± 25.4 mm | adjusting beam half cut |
| | samy (along y) | ± 25.4 mm | optimizing beam footprint |
| | samz (along z) | ± 50.8 mm | matching with beam height |
| detector vertical height adjustment | gonz | ± 75.0 mm | Along the perpendicular to the scattering plane |

Table I. Specifications of motor motions

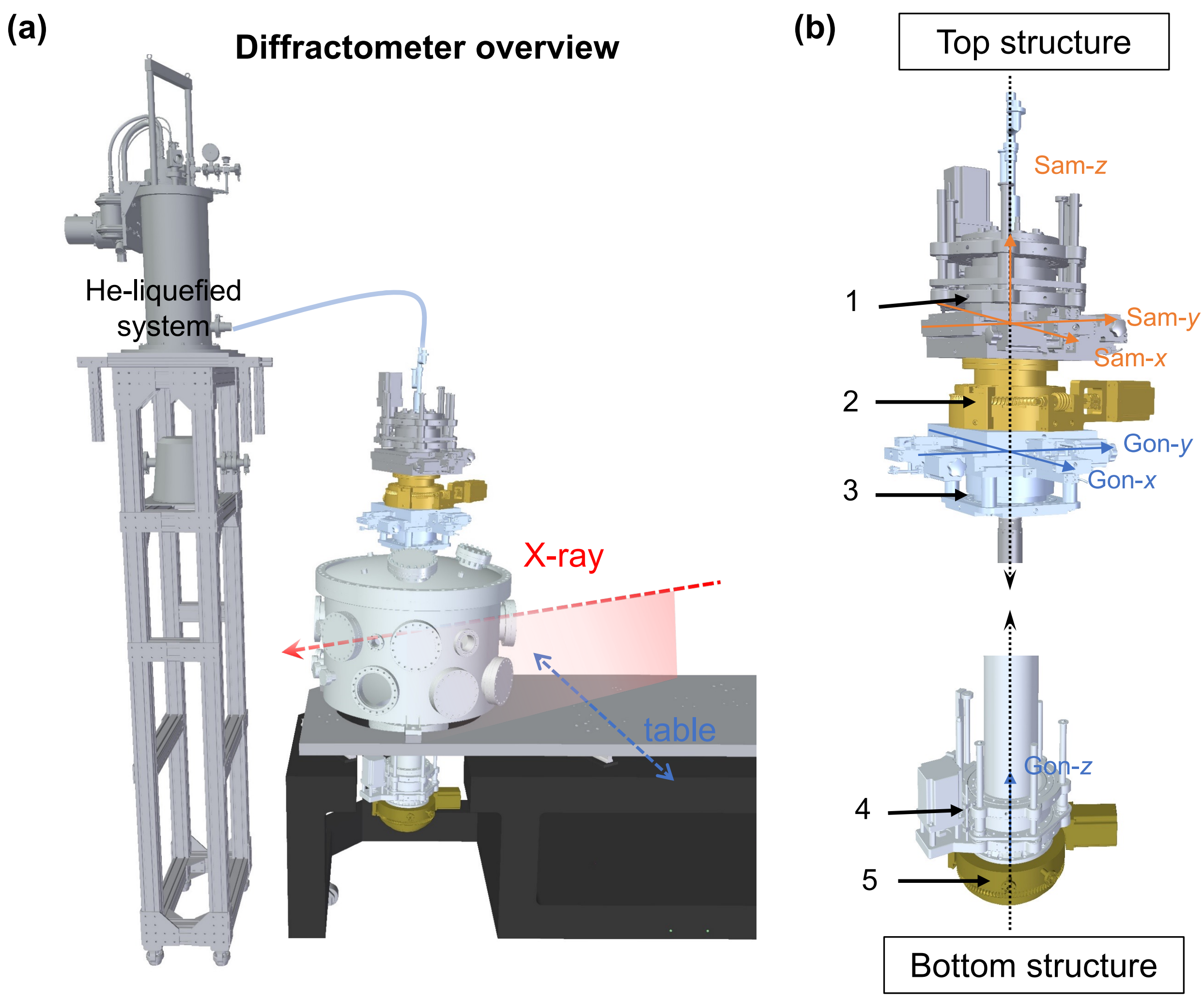

Figure 1 CAD models of (a) BL13-3 diffractometer endstation and (b) top/bottom structure assemblies. In (a) the closed-cycle helium liquefied system (RGC4) is attached in proximity to the top assembly for connection with the inlet of the sample cryostat (ST-400). In (b), 1, 2, and 3 are *x*-*y*-*z* manipulator (Sam-*x*/*y*/*z*), differentially pumped rotating platforms (DPRP), and *x*-*y* manipulator (Gon-*x*/*y*), respectively. 4 and 5 are *z* manipulator (Gon-*z*) and DPRP.

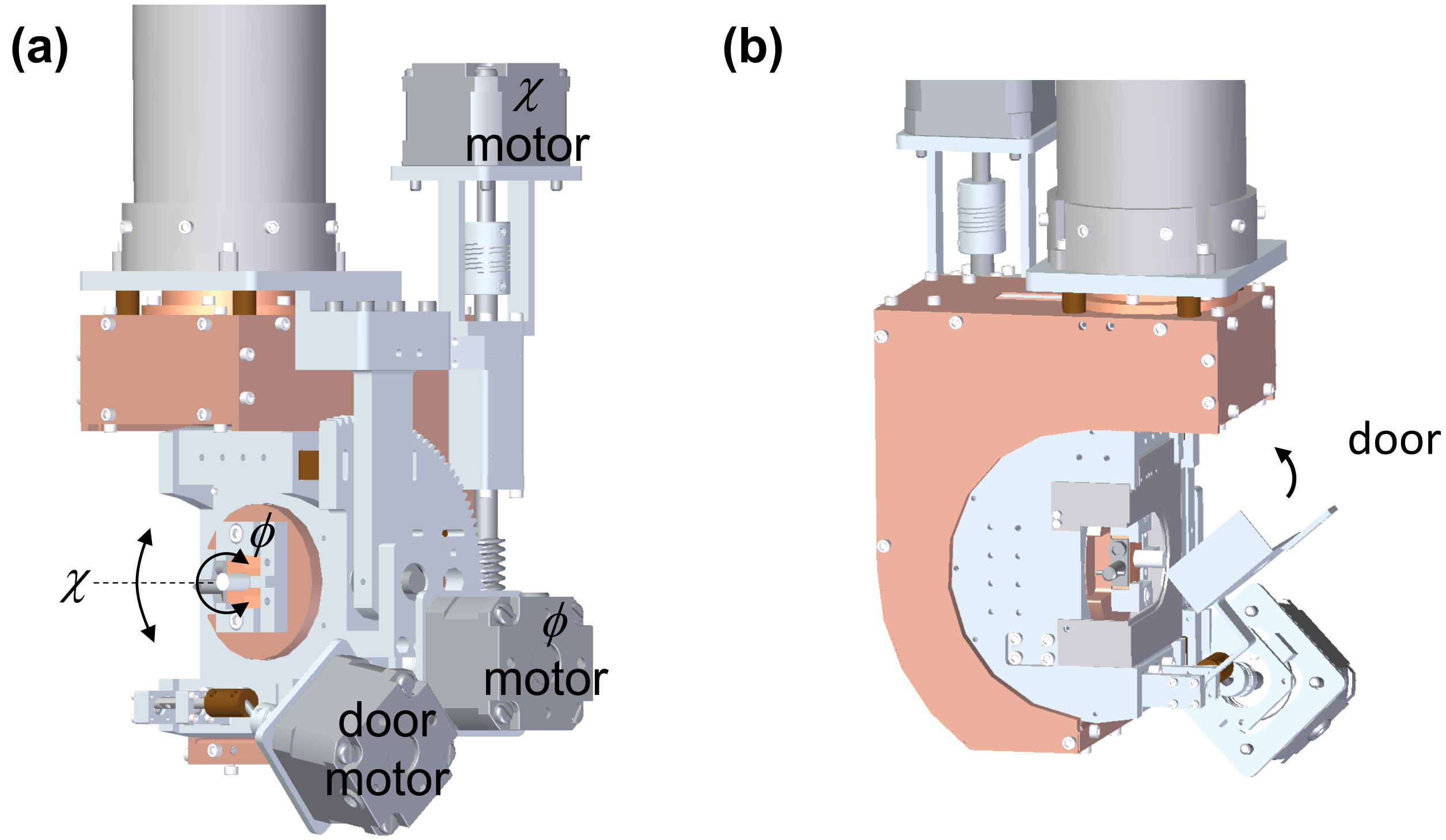

Figure 2 (a) and (b) show CAD views of the manipulator head. In (a), the purpose of the three in-vacuum motors is indicated. Two motors control ϕ and χ motions. (b) Indication of door motion by an in-vacuum motor.

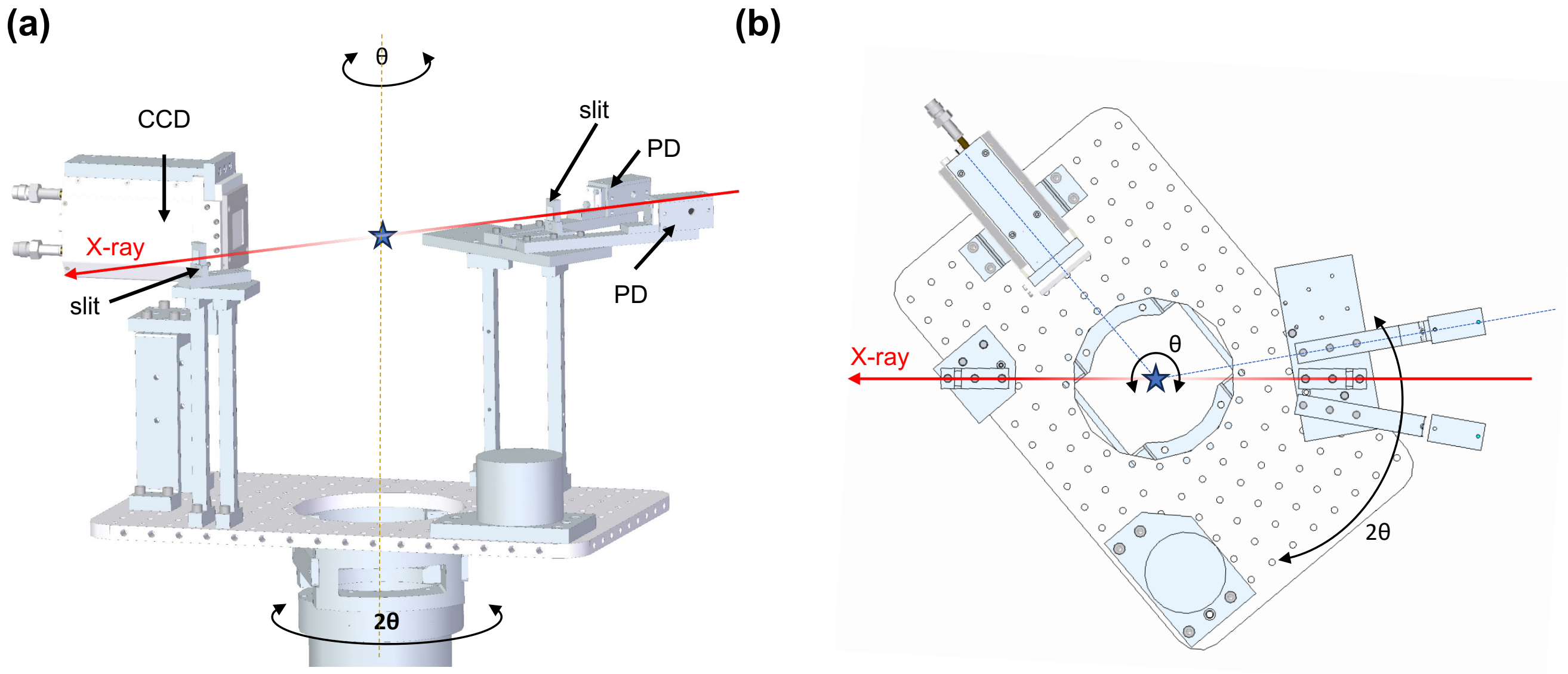

Figure 3 (a) Inside view of beamline 13-3 main chamber. (b) Top view of the experimental geometry. The sample surface is aligned to be at center of rotation (a star mark) of the main chamber.

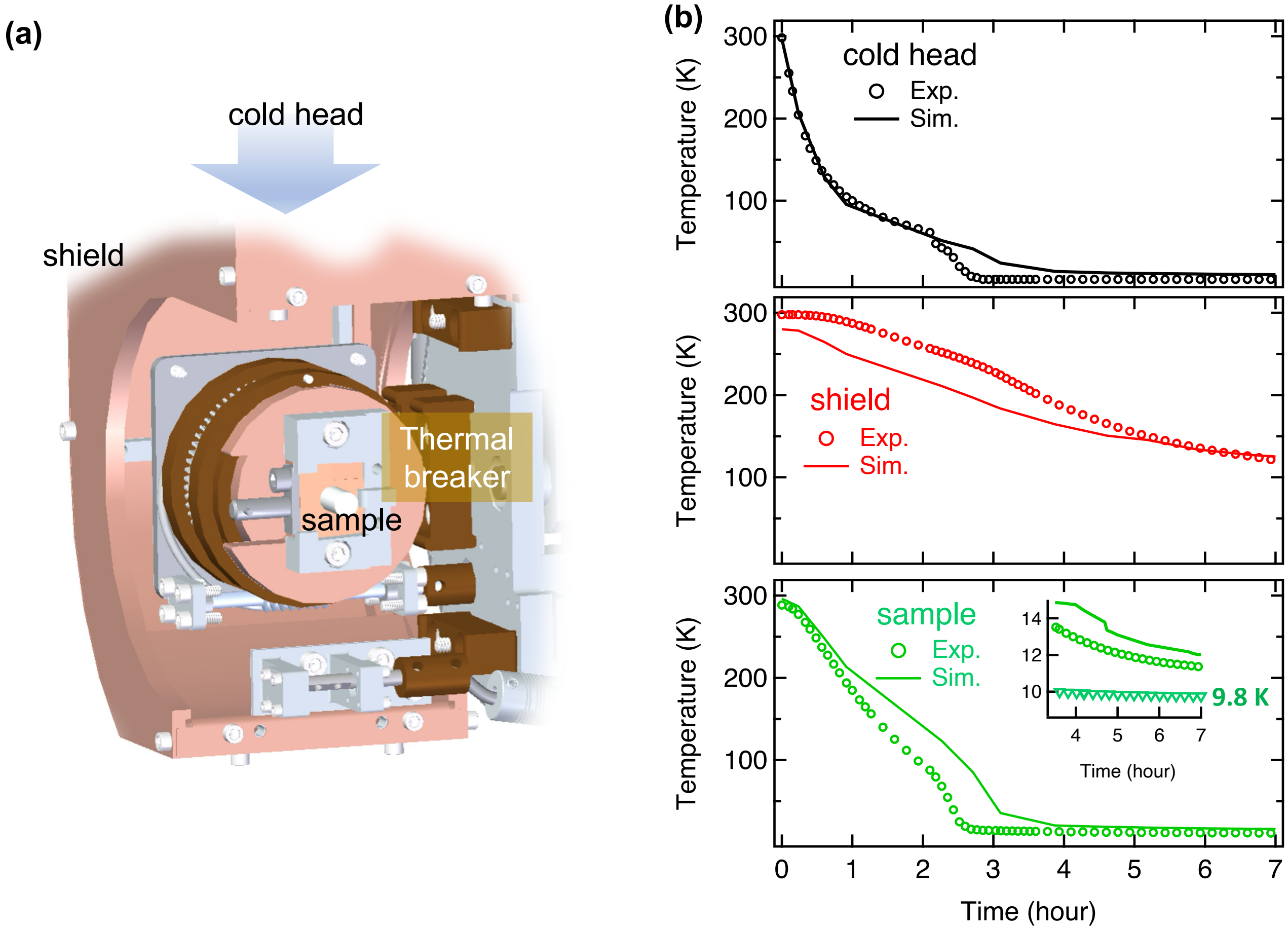

Figure 4 (a) CAD view of the sample stage. The bronze shield is hidden to show the VESPELs (brown colored) as thermal breakers around the azimuthal worm gears and other components. (b) Cooling results of CFD-FEA-simulations together with the experimental results. The colored lines are simulated data, and the open circles are experimental data. The cold head, shield, and sample are marked in black, red, and green, respectively. The inset of the bottom figure shows comparison of simulated (line), earlier experimental (open circle), and the most recent experimental (open downward triangle) results. The most recent results succeed in achieving 9.8 K.

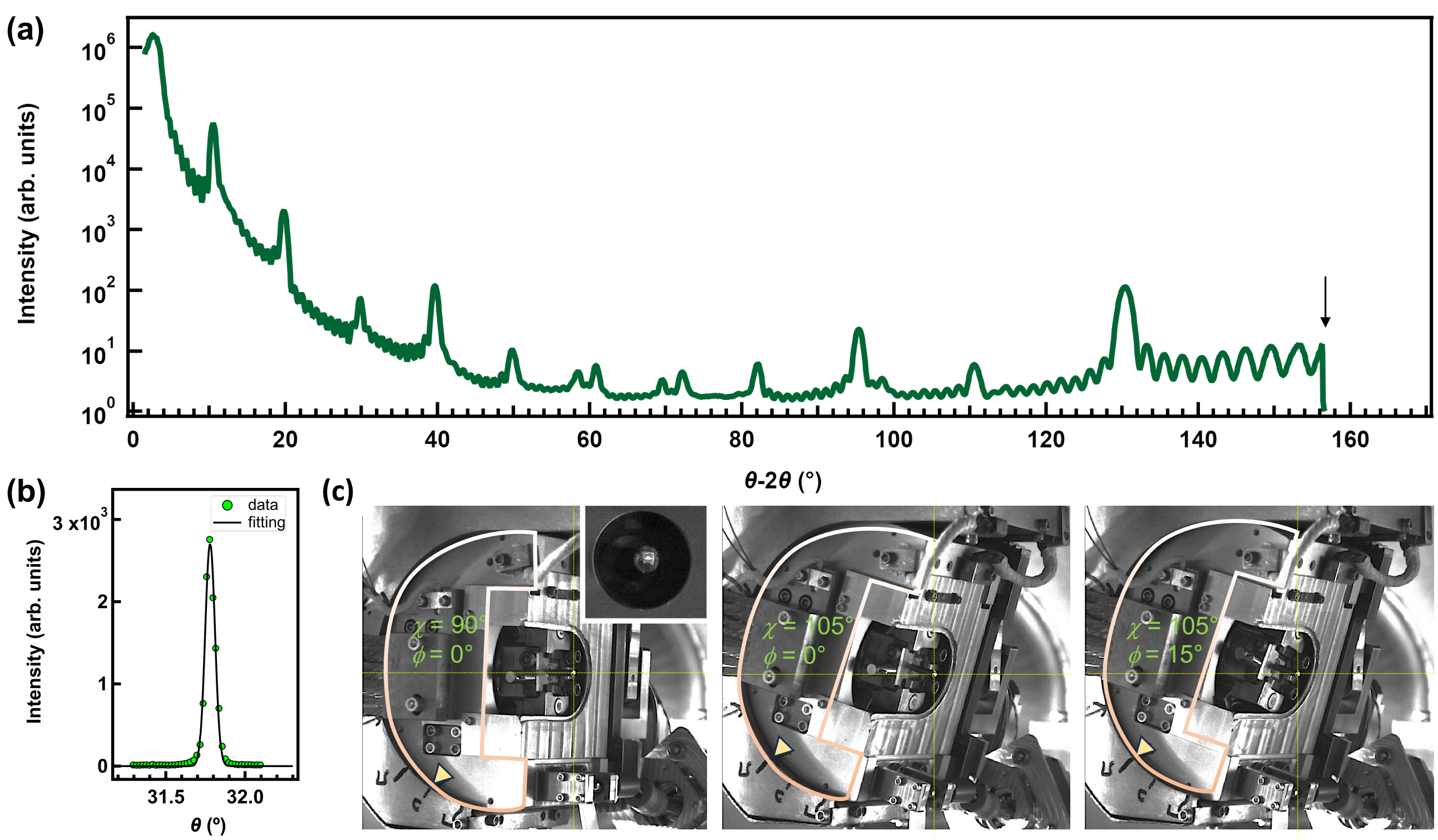

Figure 5 (a) X-ray reflectivity measurement of an SRO/STO superlattice measured using a photon energy of 1500 eV. (b) (002) Bragg reflection of a YBCO crystal using a photon energy of 1777 eV. (c) Camera snapshots of the sample stage captured at different $\chi$ and $\phi$ angles.

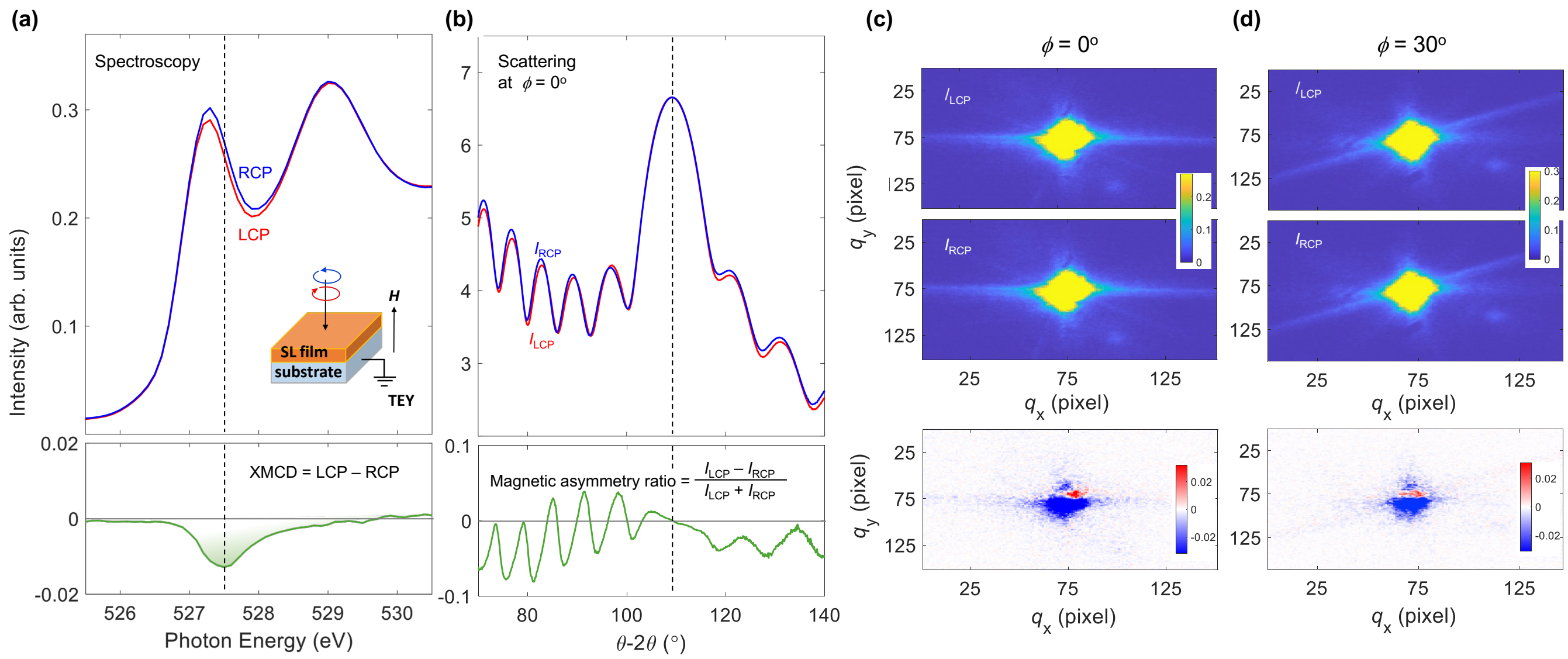

Figure 6 (a) O *K* edge XAS-TEY measurements of an SRO/STO superlattice using RCP and LCP lights and their XMCD signals. (b) Soft X-ray reflectivity measurements using RCP/LCP-polarized lights at $h\nu = 527.8$ eV and their magnetic asymmetry ratio. Magnetic asymmetry pattern collected at $h\nu = 527.8$ eV at (c) $\phi$=0° and (d) $\phi$=30°.

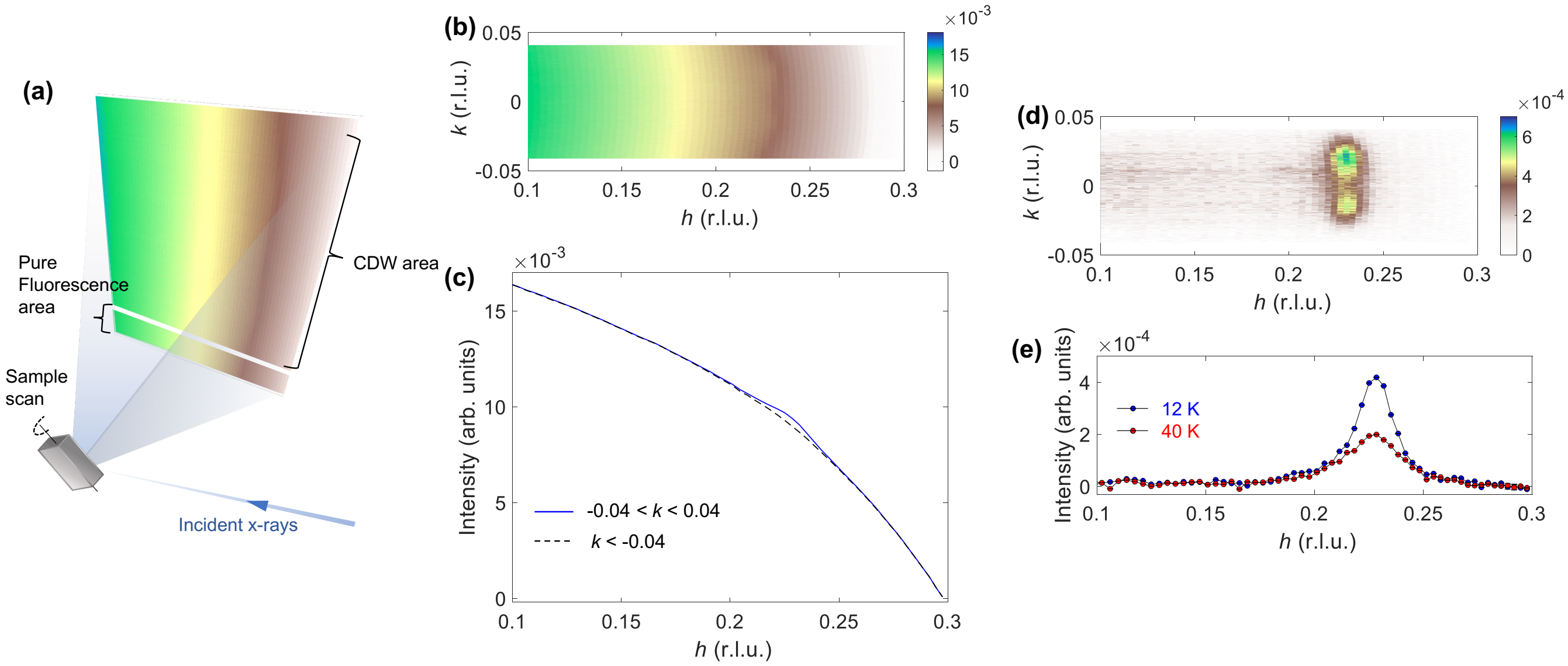

Figure 7 (a) Schematics of the RSXS measurement of LSCO (x=0.115). (b) Scattering map measured at 12 K. (c) Its projected intensity profiles along the *h*-direction with different *k* range. The pure fluorescence area ($k < -0.04$, dot line) is used for fluorescence background removal of the scattering image. (d) Scattering map after the fluorescence background removal. (e) Projected intensity profiles along the *h*-direction (after fluorescence background removal) at 12 and 40 K. The CDW peaks and profiles are clearly shown with fluorescence background removal.